\documentclass[conference]{IEEEtran}
\usepackage{cite}
\usepackage{amsmath,amssymb,amsfonts}
\usepackage{algorithmic}
\usepackage{graphicx}
\usepackage{textcomp}
\usepackage{xcolor}
\usepackage{tikz}
\usepackage{url}
\usepackage{amsthm}
\usepackage{comment}
\usepackage[linesnumbered,ruled,vlined]{algorithm2e}
\usepackage{pgfplots}

\usepackage{booktabs}
\usepackage{multirow}
\usepackage{numprint}
\npstyleenglish

\usepackage{color,soul}

\newtheorem{Remark}{Remark}
\newtheorem{Lemma}{Lemma}

\def\BibTeX{{\rm B\kern-.05em{\sc i\kern-.025em b}\kern-.08em
    T\kern-.1667em\lower.7ex\hbox{E}\kern-.125emX}}

\begin{document}
\pgfplotsset{compat=1.18}

\title{On the Security of a Code-Based PIR Scheme
}

\author{\IEEEauthorblockN{Svenja Lage}
\IEEEauthorblockA{\textit{Communications and Navigation} \\
\textit{German Aerospace Center (DLR)}\\
Oberpfaffenhofen, Germany\\
\texttt{svenja.lage@dlr.de}}
\and
\IEEEauthorblockN{Hannes Bartz}
\IEEEauthorblockA{\textit{Communications and Navigation} \\
\textit{German Aerospace Center (DLR)}\\
Oberpfaffenhofen, Germany\\
\texttt{hannes.bartz@dlr.de}}
}

\maketitle

\begin{abstract}
Private Information Retrieval (PIR) schemes allow clients to retrieve files from a database without disclosing the requested file's identity to the server. In the pursuit of post-quantum security, most recent PIR schemes rely on hard lattice problems. In contrast, the so called CB-cPIR scheme stands out as a pioneering effort to base PIR schemes on hard problems in coding theory, thereby contributing significantly to the diversification of security foundations. However, our research reveals a critical vulnerability in CB-cPIR, substantially diminishing its security levels. Moreover, a comparative analysis with state-of-the-art PIR schemes shows that CB-cPIR's advantages are reduced, making it less competitive in terms of the communication cost. Nevertheless, our findings highlight the importance of continued research into code-based PIR schemes, as they have the potential to provide a valuable alternative to lattice-based approaches.
\end{abstract}

\section{Introduction}
Private Information Retrieval (PIR) schemes, as introduced in \cite{PIR} and \cite{Kushilevitz}, enable a client to request a particular file from a database without disclosing the identity of the requested file to the server. A straightforward solution is for the user to download the entire database. Although this approach preserves the user's interest, it is clearly impractical for large databases. However, as demonstrated in \cite{PIR}, it represents the only possibility for information-theoretically secure PIR in the single-server scenario. As an alternative for single-server scenarios, various computationally secure PIR (cPIR) schemes based on computational hard problems have emerged, providing a trade-off between security and communication cost.\\ 

In the context of post-quantum cryptography, it is essential not only to consider problems that are presumed to be computationally hard on classical computers but also to develop cryptographic protocols that can withstand attacks from potential quantum computers. To address this challenge, one viable approach is to construct PIR schemes based on lattice problems, which are widely regarded as being resistant to quantum attacks. While many current PIR schemes rely on lattice problems, such as the Learning With Errors (LWE) problem or its ring variant (see \cite{SimplePIR}, \cite{LWEPIR} or \cite{XPIR}), code-based cryptography presents a promising alternative to lattice-based schemes, thereby providing a valuable diversification of cryptographic primitives. \\

One approach, which defines a PIR scheme based on hard problems in coding theory, was proposed in \cite{Original}. However, an efficient attack on this scheme was presented in \cite{Break}. In this paper, we study the code-based computationally secure CB-cPIR scheme described in \cite{Repair} and \cite{Repair1}, which is an adapted version of the original construction \cite{Original} designed to be resistant to the attack in \cite{Break}. In \cite{Repair} and \cite{Repair1}, the authors present their CB-cPIR scheme, which they thoroughly analyze for its security and compare to the state-of-the-art schemes SimplePIR \cite{SimplePIR} and XPIR \cite{XPIR}. Their evaluation suggests that the modified scheme offers improved communication and computational costs, leading them to conclude that it represents a significant step towards practical and scalable real-world PIR. Moreover, the scheme contributes to the diversification of cryptographic primitives underlying PIR schemes, which is essential for ensuring the long-term security and resilience of cryptographic systems.\\

In this paper, we present a critical attack against the CB-cPIR scheme, which significantly undermines the security levels claimed in \cite{Repair1} for the proposed parameters. Our attack, thoroughly analyzed in Section \ref{Attack}, exhibits a lower complexity than the best-known attacks prior to our work. To validate our analysis, we not only provide theoretical results but also implemented the attack, demonstrating its feasibility and effectiveness in compromising the security of the CB-cPIR scheme. While adapting the parameters can mitigate this vulnerability, our analysis in Section \ref{Comparison} reveals that the resulting rates of the CB-cPIR scheme are no longer favorable against those of XPIR. In contrast, the comparison with SimplePIR yields a more nuanced result, with the choice between the two schemes depending on the specific requirements and constraints of the particular underlying use case. Our findings correct the prevailing view on CB-cPIR's position within the landscape of computational PIR schemes. Despite this, the value of a code-based cPIR scheme remains high, and the importance of diversifying cryptographic primitives cannot be overstated.

\section{Preliminaries}\label{Preliminaries}
Let $q$ be a prime or a power of a prime. Denote the finite field of order $q$ by $\mathbb{F}_q$, and its extension field of degree $s$ by $\mathbb{F}_{q^s}$. We denote $\mathbb{F}_q^{\times}=\mathbb{F}_q\setminus\{0\}$, the set of nonzero elements in $\mathbb{F}_q$. If $\Gamma = \{\gamma_1, \ldots, \gamma_s\}$ is a basis of $\mathbb{F}_{q^s}$ over $\mathbb{F}_q$, then denote by $V = \langle \gamma_1, \ldots, \gamma_v\rangle$ for $1 \leq v \leq s$ the $v$-dimensional $\mathbb{F}_q$-linear subspace of $\mathbb{F}_{q^s}$ spanned by $\gamma_1, \ldots, \gamma_v$. Let $\Psi_{\Gamma}^V(x)$ denote the projection of $x \in \mathbb{F}_{q^s}$ onto $V$ with respect to the basis $\Gamma$.\\

For a matrix $A \in \mathbb{F}_q^{m \times n}$, denote by $A_{[i:j,:]}$ the submatrix of $A$ consisting of rows $i$ to $j$, and by $A_{[:,i:j]}$ the submatrix consisting of columns $i$ to $j$. Following the usual convention, we define the Kronecker product of 
\begin{align*}
    A=(a_{i,j})_{\substack{i=1,\ldots,m, \\j=1,\ldots,n}}\in\mathbb{F}_q^{m\times n}
\end{align*}
and $B\in\mathbb{F}_q^{v\times w}$ as 
\begin{align*}
    A\otimes B = \begin{pmatrix}
        a_{1,1}B & \hdots& a_{1,n}B \\
        \vdots && \vdots\\
        a_{m,1}B &\hdots & a_{m,n}B
    \end{pmatrix}\in\mathbb{F}_q^{mv\times nw}. 
\end{align*}

A linear code $\mathcal{C}\subseteq\mathbb{F}_{q}^n$ of length $n$ and dimension $k$ is a $k$-dimensional linear subspace of $\mathbb{F}_q^n$, denoted by $\mathcal{C} = [n, k]_q$. Every linear code $\mathcal{C}$ can be represented by a generator matrix $G \in \mathbb{F}_q^{k \times n}$ of rank $k$. Any set of $k$ linearly independent columns of $G$ forms an information set of the code.

\section{The original PIR scheme}\label{OriginalScheme}
As a counterpart to existing lattice-based PIR schemes, a code-based scheme was proposed in \cite{Original}. Although the scheme was shown to be vulnerable in \cite{Break}, we outline it here due to the close similarity between the original construction and its later modification.\\

\begin{figure}[h]
    \centering
    \begin{tikzpicture}
        \draw[thick] (0,0) rectangle (2,0.8);
        \draw[thick] (3,0) rectangle (4,0.8);
        \draw (1,0.8) -- (1,0);
        \draw (2,0.8) -- (2,0);
        \draw (3,0.8) -- (3,0);
        \node at (-0.4, 0.4) {\small $X=$};
        \node at (0.5, 0.4) {\small $X^1$};
        \node at (1.5, 0.4) {\small $X^2$};
        \node at (3.5, 0.4) {\small $X^m$};
        \node at (2.5, 0.4) {\small \ldots};
        \draw[decorate,decoration={brace,amplitude=5pt,mirror},thick] (4.1,0) -- (4.1,0.8);
        \node at (4.6,0.4) {\small $L$};
        \draw[decorate,decoration={brace,amplitude=5pt},thick] (1,0.9) -- (2,0.9);
        \node at (1.5,1.3) {\small $\delta$};
    \end{tikzpicture}
    \caption{Database construction with $m$ files $X^i$, $i=1,\ldots,m$.}
    \label{DatabaseConstruction}
\end{figure}
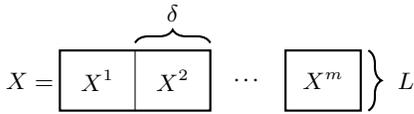

Let $v$, $s$, $n$, and $k$ be scheme parameters satisfying $v < s$ and $n < k$, and define $\delta = (s-v)(n-k)$. Suppose the database consists of $m$ files. Each file $X^i$ is represented as a matrix $X^i \in \mathbb{F}_q^{L \times \delta}$, where $L$ is chosen sufficiently large to accommodate all files. Consequently, the entire database is represented as $X \in \mathbb{F}_q^{L \times m\delta}$, as illustrated in Figure \ref{DatabaseConstruction}.

\vspace{-0.3cm}

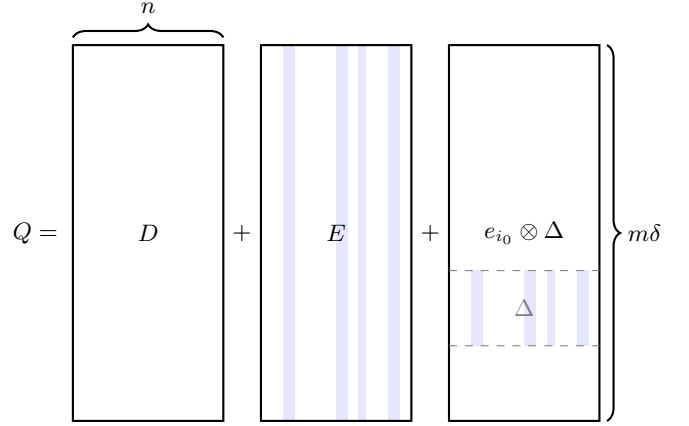
\begin{figure}[h]
    \centering
    \begin{tikzpicture}

        \fill[blue!10](2.8,0) rectangle (2.95,5);
        \fill[blue!10](3.5,0) rectangle (3.65,5);
        \fill[blue!10](3.8,0) rectangle (3.9,5);
        \fill[blue!10](4.2,0) rectangle (4.35,5);

        \fill[blue!10](5.3,1) rectangle (4.95+0.5,2);
        \fill[blue!10](6,1) rectangle (5.65+0.5,2);
        \fill[blue!10](5.8+0.5,1) rectangle (5.9+0.5,2);
        \fill[blue!10](6.2+0.5,1) rectangle (6.35+0.5,2);

        \draw[thick] (0,0) rectangle (2,5);
        \draw[thick] (2.5,0) rectangle (4.5,5);
        \draw[thick] (5,0) rectangle (7,5);

        \node at (-0.5,2.5) {\small $Q=$};
        \node at (1,2.5) {\small $D$};
        \node at (3.5,2.5) {\small $E$};
        \node at (6,2.5) {\small $e_{i_0}\otimes\Delta$};
        \node[text=gray] at (6,1.5) {\small $\Delta$};
        \node at (2.25,2.5) {\small $+$};
        \node at (4.75,2.5) {\small $+$};

        \draw[dashed,gray] (5,1) -- (7,1);
        \draw[dashed,gray] (5,2) -- (7,2);

        \draw[decorate,decoration={brace,amplitude=5pt},thick] (0,5.1) -- (2,5.1);
        \node at (1,5.5) {\small $n$};
        \draw[decorate,decoration={brace,amplitude=5pt,mirror},thick] (7.1,0) -- (7.1,5);
        \node at (7.6,2.5) {\small $m\delta$};
    \end{tikzpicture}
    \caption{Query construction in the original code-based PIR scheme \cite{Original}. }
    \label{QueryOriginal}
\end{figure}

\textbf{Query generation:}
To request file $X^{i_0}$ for some $i_0\in\{1,\ldots,m\}$, the user performs the following steps:
\begin{itemize}
    \item Select a random $[n, k]_{q^s}$-linear code $\mathcal{C}$. 
    \item Construct a matrix $D \in \mathbb{F}_{q^s}^{m\delta \times n}$, where each row is a codeword from $\mathcal{C}$. Randomly choose an information set $I \subset \{1, \ldots, n\}$ with $|I| = k$.
    \item Choose a random basis $\Gamma = \{\gamma_1, \gamma_2, \ldots, \gamma_s\}$ of $\mathbb{F}_{q^s}$ over $\mathbb{F}_q$. Let $V = \langle \gamma_1, \ldots, \gamma_v \rangle $ be the $\mathbb{F}_q$-linear, $v$-dimensional subspace of $\mathbb{F}_{q^s}$ spanned by the first $v$ basis vectors, and let $W = \langle \gamma_{v+1}, \ldots, \gamma_s\rangle$.
    \item Randomly choose $\tilde{E} \in V^{m\delta \times (n-k)}$ and extend it to a matrix $E \in V^{m\delta \times n}$ by inserting zero-columns at the positions of the information set $I$. 
    \item Randomly choose $\tilde{\Delta} \in W^{\delta \times (n-k)}$ with full row-rank over $\mathbb{F}_q$, i.e., $rk_{\mathbb{F}_q}(\tilde{\Delta}) = \delta$, and extend $\tilde{\Delta}$ to $\Delta \in W^{\delta \times n}$ by inserting zero-columns at the positions of the information set $I$.
\end{itemize}
The final request is computed as
\begin{align*}
    Q = D + E + e_{i_0} \otimes \Delta, 
\end{align*}
where $e_{i_0}\in\mathbb{F}_{q^s}^{m\times 1}$ is the $i_0$-th unit vector. A graphical representation of the query generation is shown in Figure \ref{QueryOriginal}. 

\textbf{Answer: }
Upon receiving $Q \in \mathbb{F}_{q^s}^{m \delta \times n}$, the server computes the response as
\begin{align*}
    R = X \cdot Q \in \mathbb{F}_{q^s}^{L \times n},
\end{align*}
and returns the matrix $R$ to the client.\\

\textbf{File extraction:}
To recover the desired file $X^{i_0}$, the user examines the $j$-th row of $R$, denoted as $R_j$ for $j = 1, \ldots, L$. The $j$-th row of $R$ is given by
\begin{align*}
    R_j &= X_{[j,:]} \cdot Q \\
    &= \sum_{l=1}^m X_{[j,(l-1)\delta+1:l\delta]} Q_{[(l-1)\delta+1:l\delta,:]} \\
    &= \sum_{l=1}^m X_{[j,:]}^l Q_{[(l-1)\delta+1:l\delta,:]},
\end{align*}
using a decomposition of $Q$ into blocks of length $\delta$ such that the files $X^1, \ldots, X^m$ can be considered separately. Inserting the construction of $Q$ yields
\begin{align*}
    R_j=  \sum_{l=1}^m X_{[j,:]}^l (D_{[(l-1)\delta+1:l\delta,:]} + E_{[(l-1)\delta+1:l\delta,:]}+&\\
     (e_{i_0}\otimes \Delta)_{[(l-1)\delta+1:l\delta,:]}).& 
\end{align*}
However, $e_{i_0} \otimes \Delta$ is zero in all rows except for the $i_0$-th block, such that 
\begin{align}\label{Aj}
    R_j = \underbrace{X_{[j,:]}^{i_0}\Delta}_{\text{zero at positions in }I}   + \underbrace{\sum_{l=1}^m X_{[j,:]}^l D_{[(l-1)\delta+1:l\delta,:]}}_{=:A_j} +\notag \\
 \underbrace{ \sum_{l=1}^m X_{[j,:]}^l E_{[(l-1)\delta+1:l\delta,:]}}_{\text{zero at positions in }I}. 
\end{align}
It is worth noting that since the rows of $D$ are codewords, $A_j$ itself is a codeword. Since the other two terms have support only outside the information set $I$, the client can readily recover $A_j$ from $R_j$. By subtracting the codeword from the row of the answer, we obtain
\begin{align*}
    R_j-A_j = \underbrace{X_{[j,:]}^{i_0}\Delta}_{\in W^{1\times n}} + \underbrace{\sum_{l=1}^m X_{[j,:]}^l E_{[(l-1)\delta+1:l\delta,:]}}_{\in V^{1\times n}}.
\end{align*}
Applying the projection $\Psi_{\Gamma}^W$ to $R_j - A_j$ results in $X_{[j,:]}^{i_0}\Delta$, where $\Delta$ is a full row-rank matrix. Consequently, the client can recover $X_{[j,:]}^{i_0}$. Repeating this process for all rows $R_j$ with $j = 1, \ldots, L$, the entire file $X^{i_0}$ is recovered.\\

The authors presented various parameter sets. The field order $q$ takes values in $\{16,32,64\}$, while the extension degree $s$ is consistently set to $s=32$. The subspace dimension $v$ varies between $v=16$ and $v=31$, and the code parameters $n$ and $k$ are always given as $n = 100$ and $k = 50$.

\section{The Subquery attack}
In \cite{Break}, Bordage and Lavazelle introduced an attack on the scheme that breaks it for all reasonable parameter choices, provided the database is not too small. They observed that the $\mathbb{F}_q$-rank of $D + E$ is significantly smaller than the rank of the full query matrix $Q = D + E + e_{i_0} \otimes \Delta$ and utilized this observation to develop an efficient attack. Informally, the attack proceeds by iteratively removing individual blocks of size $\delta$ from $Q$ and observing the resulting deviation in rank. \\

More formally, let $Q[j]$ denote the submatrix of $Q$ obtained by deleting rows $[j\delta+1: (j+1)\delta]$. The authors of \cite{Break} demonstrated that, for $j = i_0$, removing the only $\Delta$-dependent component in $Q$, results in
$rk_{\mathbb{F}_q}(Q[i_0]) \leq sn - \delta$. In contrast, for $j\neq i_0$, it is highly unlikely that the $\mathbb{F}_q$-rank of $Q[j]$ for $j \neq i_0$ is this small. In fact, according to \cite[Proposition 2]{Break}, the probability is at most
\begin{align*}
    P(rk_{\mathbb{F}_q}(Q[j])\leq sn-\delta) \leq \genfrac{[}{]}{0pt}{}{sn-\delta}{sn-2\delta}_q q^{-\delta^2(m-1)}, 
\end{align*}
where $\genfrac{[}{]}{0pt}{}{a}{b}_q$ represents the $q$-binomial coefficient \cite{qBinomial} for $a \geq b$. Therefore, by examining the ranks of the submatrices $Q[1], \ldots, Q[m]$, the index of interest $i_0$ can be identified with high probability. As a result, the authors were able to break the scheme within minutes on a standard computer for all parameter sets proposed in \cite{Original}.\\

To observe differences in rank, it is required that 
\begin{align*}
    (m-1)\delta \geq n,
\end{align*}
ensuring that the number of rows in $Q[j]$ exceeds the number of columns. This implies a necessary condition on $m$, namely $m\geq \frac{n}{\delta}+1$. Increasing $m$ beyond this threshold improves the probability of successfully distinguishing the ranks. \\

\section{The repair}
In 2024, Verma and Hollanti proposed a modified version for this PIR scheme (see \cite{Repair} and its full version \cite{Repair1}), which remains structurally close to the the original construction but is designed to resist the subquery attack described in the previous section. The central idea of their work is to distribute $\Delta$ across all blocks of $Q$, while introducing a small perturbation in block $i_0$ to conceal the index of interest. \\

The revised protocol, called CB-cPIR, operates in the following manner. Consider again the scenario in which a user seeks to retrieve a single file $X^{i_0}$ from the server without revealing any information regarding the requested file. The database is constituted as described in Section \ref{OriginalScheme}. The parameters $q$, $s$, $v$, as well as $n$ and $k$, are scheme parameters that retain their previously defined meanings. \\

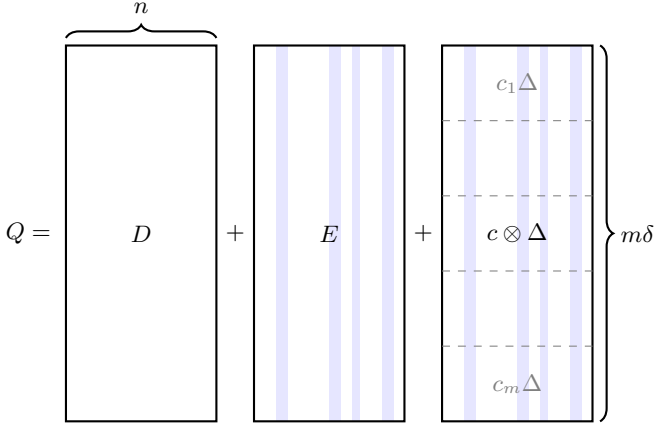
\begin{figure}[h]
    \centering
    \begin{tikzpicture}

        \fill[blue!10](5.3,0) rectangle (4.95+0.5,5);
        \fill[blue!10](6,0) rectangle (5.65+0.5,5);
        \fill[blue!10](5.8+0.5,0) rectangle (5.9+0.5,5);
        \fill[blue!10](6.2+0.5,0) rectangle (6.35+0.5,5);
        
        \fill[blue!10](2.8,0) rectangle (2.95,5);
        \fill[blue!10](3.5,0) rectangle (3.65,5);
        \fill[blue!10](3.8,0) rectangle (3.9,5);
        \fill[blue!10](4.2,0) rectangle (4.35,5);

        \draw[thick] (0,0) rectangle (2,5);
        \draw[thick] (2.5,0) rectangle (4.5,5);
        \draw[thick] (5,0) rectangle (7,5);

        \node at (-0.5,2.5) {\small $Q=$};
        \node at (1,2.5) {\small $D$};
        \node at (3.5,2.5) {\small $E$};
        \node at (6,2.5) {\small $c\otimes\Delta$};
        \node[text=gray] at (6,4.5) {\small $c_1\Delta$};
        \node[text=gray] at (6,0.5) {\small $c_m\Delta$};

        \node at (2.25,2.5) {\small $+$};
        \node at (4.75,2.5) {\small $+$};

        \draw[dashed,gray] (5,1) -- (7,1);
        \draw[dashed,gray] (5,2) -- (7,2);
        \draw[dashed,gray] (5,3) -- (7,3);
        \draw[dashed,gray] (5,4) -- (7,4);

        \draw[decorate,decoration={brace,amplitude=5pt},thick] (0,5.1) -- (2,5.1);
        \node at (1,5.5) {\small $n$};
        \draw[decorate,decoration={brace,amplitude=5pt,mirror},thick] (7.1,0) -- (7.1,5);
        \node at (7.6,2.5) {\small $m\delta$};
    \end{tikzpicture}
    \caption{Construction of $Q$ as part of the query in the repaired scheme as described in \cite{Repair}. }
    \label{ModifiedQuery}
\end{figure}

\textbf{Query generation: }
\begin{itemize}
    \item Sample uniformly at random a vector $\beta \in (\mathbb{F}_q^{\times})^m$ and let $c = e_{i_0}+\beta$, where $e_{i_0}\in\mathbb{F}_{q^s}^m$ is the $i_0$-th unit vector. 
    \item Choose a linear code $\mathcal{C}$ with an information set $I$ and define subspace $V$ and $W$ as in the original scheme to construct matrices $D$, $E$ and $\Delta$.
    \item As a first part of the query, let $Q = D + E + c \otimes \Delta$. 
    \item Choose another linear code $\mathcal{C}_{\beta}$ with an information set $I_{\beta}$ and define subspaces $V_{\beta}$ and $W_{\beta}$ to generate matrices $D_{\beta}$, $E_{\beta}$ and $\Delta_{\beta}$ according to the same procedure. 
    \item Set $Q_{\beta} = D_{\beta}+E_{\beta} +\beta \otimes \Delta_{\beta}$. 
\end{itemize}
The user sends $Q$ and $Q_{\beta}$ to the server. An illustration of the query generation in the modified scheme is shown in Figure \ref{ModifiedQuery}. \\

It is important to note that the vector $c$ contains at least $(m-1)$ non-zero entries. More specifically, $c$ can be chosen to have no zero entries, ensuring that the Kronecker product $c\otimes \Delta$ affects every block of $Q$. However, the introduction of $c$ incurs additional cost: an additional matrix $Q_{\beta}$ must be computed and transmitted, which subsequently affects the overall PIR rate.\\

\textbf{Answer: } Following the methodology of the original scheme, the server computes the product $R = X \cdot Q$. Additionally, it computes the product $R_{\beta} = X \cdot Q_{\beta}$ and transmits both matrices to the client.\\

\textbf{File extraction:}
As in the original scheme, the client selects a row $R_j$ of $R$, where $j \in \{1, \ldots, L\}$, from the matrix $R$. Using the definition of $A_j$ in (\ref{Aj}), the client computes and removes the codeword component in $R$ as 
\begin{align*}
    R_j-A_j = \underbrace{\sum_{l=1}^m X_{[j,:]}^l E_{[(l-1)\delta+1:l\delta,:]}}_{\in V^{1\times n}} + \underbrace{X_{[j,:]}(c\otimes \Delta)}_{\in W^{1\times n}}. 
\end{align*}
Unlike the original setting, the $\Delta$-dependent part of the response now involves the entire database, rather than being limited to the file $X^{i_0}$. To isolate the desired contribution, the client projects the resulting vector onto the subspace $W$, obtaining $X_{[j,:]}(c\otimes \Delta)$. Owing to the full row-rank property of $\Delta$, the client constructs
\begin{align*}
    X_{[j,:]}(c \otimes I_{\delta\times\delta}) = X_{[j,:]}((e_{i_0}+\beta)  \otimes I_{\delta\times\delta}). 
\end{align*}
Proceeding in a similar manner with $Q^{\beta}$, the client observes $X_{[j,:]}(\beta \otimes I_{\delta \times \delta})$. Subtracting both terms, the client is capable of reconstructing $X^{i_0}_{[j,:]}$. Repeating this process for all rows ultimately yields the desired file.\\

The authors of \cite{Repair} and \cite{Repair1} further proposes a modification to enhance the rate of the scheme. This modification allows for $f$ files to be requested simultaneously using the same vector $\beta$, thereby enabling the client to store $R_{\beta}$ at the client's side in the first query. Subsequent queries no longer require the transmission of $Q_{\beta}$, thereby improving the overall rate. As stated in \cite[Theorem 2]{Repair}, with $X^{i_0}\in\mathbb{F}_q^{L\times \delta}$, $Q,Q_{\beta}\in\mathbb{F}_{q^s}^{m\delta \times n}$ as well as $R,R_{\beta}\in \mathbb{F}_{q^s}^{L\times n}$, the PIR rate is given by
\begin{align*}
    R_{\text{CB-cPIR}} = \frac{fL\delta\log_2(q)}{(f+1)(m\delta n +Ln)\log_2(q^s)}.
\end{align*}
In scenarios where multiple files are requested concurrently, the weight of $c$ can be reduced by the number of files requested. This reduction in weight has the potential to introduce a vulnerability, particularly for large values of $f$. However, it has been countered by the authors that the construction of $c$ can always be performed in such a manner that it possesses a sufficiently high weight, thereby rendering this attack infeasible. \\

A thorough analysis of various attacks, including information set decoding, subspace attack, original subquery attack, and an advanced variant, was conducted in \cite{Repair1}. This led to the proposal of parameters with corresponding rates and security levels, which are summarized in Table \ref{ParamRate} and Table \ref{AttackCom}. Note that the displayed rate is approximated for the scenario where $L \gg m\delta$ and $f=1$, resulting in $R_{\text{CB-cPIR}} \approx \frac{\delta}{2ns}$. Furthermore, an iterative version of the scheme was introduced, which reorders the database into a square or hypercube structure, inspired by \cite{Kushilevitz}, to optimize communication costs. This approach is particularly effective when dealing with small file sizes in large databases. In comparison to the recent lattice-based protocols SimplePIR \cite{SimplePIR} and XPIR \cite{XPIR}, the authors claim lower communication and computational costs. 

\begin{table}
    \centering
    \caption{Proposed parameters and resulting approximative rate in the case $L\gg m\delta$ and $f=1$ for the modified PIR scheme in \cite{Repair1}.}
    \begin{tabular}{@{}lrrrr|r@{}}
    \toprule
        q & s & v & n & k & $R_{\text{CB-cPIR}}$ \\
    \midrule
    \midrule
    32 & 32& 31 & 100 & 50 & 1/128 \\
    32 & 32 & 30 & 100 & 50& 1/64\\
    $2^{16}$ & 12 & 10 & 100 & 50& 1/24\\
    $2^{32}-5$ & 6 & 4 & 120 & 60& 1/12\\
    $2^{32}$ & 5 & 3 & 100 & 50& 1/10\\
    $2^{61}-1$ & 6 & 2 & 100 & 50 & 1/6 \\
     \bottomrule
    \end{tabular}
    \label{ParamRate}
\end{table}

\section{An attack on the repair}\label{Attack}
We will now demonstrate how to compromise the revised scheme. To achieve this, we will divide the matrix $Q$ into blocks consisting of $\delta$ rows each as
\begin{align*}
    Q= \begin{pmatrix}
        Q^1 \\
        \vdots\\
        Q^m 
    \end{pmatrix} = \begin{pmatrix}
        D^1 \\
        \vdots\\
        D^m 
    \end{pmatrix} + \begin{pmatrix}
        E^1 \\
        \vdots\\
        E^m 
    \end{pmatrix} + \begin{pmatrix}
        c_1 \Delta \\
        \vdots\\
        c_m \Delta 
    \end{pmatrix},
\end{align*}
where 
\begin{align}\label{c}
    c_i = \begin{cases}
        \beta_i &\text{ if }i\neq i_0\\
        1+\beta_{i_0} & \text{ if }i=i_0. 
    \end{cases}
\end{align}
The attack will be carried out in two consecutive steps.  \\

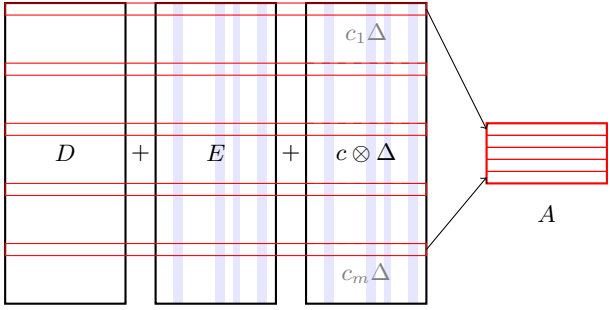
\begin{figure}[h]
    \centering
    \begin{tikzpicture}[scale=0.8]

        \fill[blue!10](5.3,0) rectangle (4.95+0.5,5);
        \fill[blue!10](6,0) rectangle (5.65+0.5,5);
        \fill[blue!10](5.8+0.5,0) rectangle (5.9+0.5,5);
        \fill[blue!10](6.2+0.5,0) rectangle (6.35+0.5,5);
        
        \fill[blue!10](2.8,0) rectangle (2.95,5);
        \fill[blue!10](3.5,0) rectangle (3.65,5);
        \fill[blue!10](3.8,0) rectangle (3.9,5);
        \fill[blue!10](4.2,0) rectangle (4.35,5);

        \draw[thick] (0,0) rectangle (2,5);
        \draw[thick] (2.5,0) rectangle (4.5,5);
        \draw[thick] (5,0) rectangle (7,5);

        \node at (1,2.5) {\small $D$};
        \node at (3.5,2.5) {\small $E$};
        \node at (6,2.5) {\small $c\otimes\Delta$};
        \node[text=gray] at (6,4.5) {\small $c_1\Delta$};
        \node[text=gray] at (6,0.5) {\small $c_m\Delta$};

        \node at (2.25,2.5) {\small $+$};
        \node at (4.75,2.5) {\small $+$};

        \draw[dashed,gray] (5,1) -- (7,1);
        \draw[dashed,gray] (5,2) -- (7,2);
        \draw[dashed,gray] (5,3) -- (7,3);
        \draw[dashed,gray] (5,4) -- (7,4);

        \draw[red] (0,5) rectangle (7,4.8);
        \draw[red] (0,4) rectangle (7,3.8);
        \draw[red] (0,3) rectangle (7,2.8);
        \draw[red] (0,2) rectangle (7,1.8);
        \draw[red] (0,1) rectangle (7,0.8);

        \draw[thick,red] (8,2) rectangle (10,3);
        \draw[red] (8,2) rectangle (10,2.2);
        \draw[red] (8,2.2) rectangle (10,2.4);
        \draw[red] (8,2.4) rectangle (10,2.6);
        \draw[red] (8,2.6) rectangle (10,2.8);
        \draw[red] (8,2.8) rectangle (10,3);
        \node at (9,1.5) {\small $A$};

        \draw[->] (7,4.9) -- (8,2.9);
        \draw[->] (7,0.9) -- (8,2.1);

    \end{tikzpicture}
    \caption{Visualization of the fist step within the attack: Construction of the auxiliary matrix $A$. }
    \label{ConstructA}
\end{figure}

\textbf{Step 1: Constructing an Auxiliary Matrix} \\
The first step is an auxiliary step that will enable us to distinguish between blocks with and without a $\Delta$-dependent part later on. To this end, we construct a matrix $A \in \mathbb{F}_{q^s}^{(ns-\delta+1) \times n}$ with a similar structure to $Q$, but with low rank. Assume that the number of files $m$ in the database fulfills $m > ns - \delta$. Then $A$ consists of the first row of each of the matrices $Q^1, \ldots, Q^{ns-\delta+1}$ as displayed in Figure \ref{ConstructA}. Observe that $A$ shares a similar structure with $Q$, as each row comprises a codeword from the linear code plus an error term from $V$ in the columns outside the information set. However, unlike $Q$, every row of $A$ features the same vector from $\Delta$, albeit scaled by different factors from $\mathbb{F}_q$. Hence, following the same arguments as in \cite[Corollary 3.2]{Break}, we have $rk_{\mathbb{F}_q}(A)\leq ns-\delta+1$.\\

\textbf{Step 2: Identify the index of interest} \\
We now proceed to analyze each block $Q^1, \ldots, Q^m$ individually with the objective of reconstructing the index of interest $i_0$. Let $J = \{1, \ldots, m\}$ be the set of indices that are candidates for the index of interest. We randomly select two indices $i, j \in J$ and apply the following procedure to determine whether one of these indices coincides with $i_0$.\\

\textbf{Step 2.1: Identify }$\alpha\in\mathbb{F}_q$\textbf{ with }
\begin{align}\label{cond}
   \alpha c_i + c_j = 0. 
\end{align}
To this end, consider the matrix
\begin{align*}
    D^{i,j} = D^{i,j}(\alpha_2,\ldots,\alpha_{\delta}) = \begin{pmatrix}
        \alpha_2 Q^i_{[2,:]}  + Q^j_{[2,:]}\\
        \vdots\\
        \alpha_{\delta}Q^i_{[\delta,:]}    + Q^j_{[\delta,:]}
    \end{pmatrix}
\end{align*}
for different values $\alpha_2,\ldots,\alpha_{\delta} \in \mathbb{F}_q$. Note that if the equation 
\begin{align*}
   \alpha_k c_i + c_j = 0 
\end{align*}
holds in $\mathbb{F}_q$ for some $k\in\{2,\ldots,\delta\}$, then the $k$-th row of $D^{i,j}$ will not have a $\Delta$-dependent component. However, we cannot determine this solely by analyzing the rank of $D^{i,j}$, as the matrix has only $(\delta-1)$ rows, which would dictate its rank regardless of the presence of a $\Delta$-dependent component. Instead, we can verify whether condition (\ref{cond}) holds for one $\alpha_k$, $k\in\{2,\ldots,\delta\}$, by leveraging the following observation. If a row of $D^{i,j}$ does not have a $\Delta$-dependent component, then appending this row to $A$ will not increase the rank of $A$. On the other hand, if the row does have a $\Delta$-dependent component, then appending it to $A$ will increase the rank by one. We can efficiently test all values of $\alpha_2,\ldots,\alpha_{\delta}$ simultaneously by simply appending $D^{i,j}$ to $A$. If the resulting matrix has rank $rk_{\mathbb{F}_q}(A)+\delta-1$, then none of the tested values satisfy condition (\ref{cond}). Conversely, if the rank is less than $rk_{\mathbb{F}_q}(A) + \delta - 1$, then, given that $\Delta$ has full rank by definition, there is one $\alpha\in\{\alpha_2,\ldots,\alpha_{\delta}\}$ fulfilling (\ref{cond}). \\

In the worst-case scenario, we need to calculate the rank of an $\mathbb{F}_q^{ns \times ns}$ matrix $\lceil\frac{q}{\delta - 1}\rceil$ times to identify the tuple $(\alpha_2, \ldots, \alpha_{\delta})$ that contains the correct value of $\alpha$. To pinpoint this value from the tuple, we can employ a binary-search-like algorithm, {as introduced in \cite{lage2025cryptanalysislatticebasedpirscheme} for an attack on a different PIR scheme. Specifically, we can append only the first half of the rows of $D^{i,j}$ to $A$ and check whether the rank increases by the number of rows added. If it does not, the correct value of $\alpha$ is contained in the first half; otherwise, it is contained in the second half. Consequently, we can eliminate half of the possible values of $\alpha$ with a single rank calculation. By iteratively applying this procedure, we can identify the sought-after value of $\alpha$ from the right tuple in at most $\log(\delta)$ rank calculations.\\

\textbf{Step 2.2: Check if }$i_0\in\{i,j\}$\\
Determining the value of $\alpha \in \mathbb{F}_q$ with 
\begin{align*}
    0=\alpha c_i+c_j=\begin{cases}
        \alpha\beta_i + \beta_j  & \text{ if } i_0\neq i \text{ and } i_0\neq j\\
        \alpha\beta_i + 1+ \beta_{i_0} & \text{ if } i_0=j\\
        \alpha(1+\beta_{i_0}) +\beta_j & \text{ if } i_0=i
    \end{cases}
\end{align*}
does not directly yield $c_i$ or $c_j$. Nevertheless, the relation between $c_i$ and $c_j$ can be utilized to verify whether the index of interest is equal to $i$ or $j$. To achieve this, we repeat Step 1 with $Q_{\beta}$ instead of $Q$ and obtain a matrix $A_{\beta}\in\mathbb{F}_{q^s}^{(ns-\delta+1)\times n}$ with $rk_{\mathbb{F}_q}(A_{\beta})=ns-\delta+1$. Subsequently, we append the vector 
\begin{align*}
    \alpha {Q_{\beta}^i}_{[2,:]}-{Q_{\beta}^j}_{[2,:]}
\end{align*}
 to $A_{\beta}$. If the rank of the resulting matrix increases by one, then it follows that
\begin{align*}
    \alpha\beta_i +\beta_j\neq 0, 
\end{align*}
which implies $i_0\in\{i,j\}$. In this case, repeat Step 2 with the same value of $i$ but a different value of $j$ to check if $i_0=i$ or $i_0=j$. Conversely, if the rank does not increase, then $\alpha\beta_i +\beta_j= 0$, which implies that $i_0\neq i$ and $i_0\neq j$. We can thus exclude these indices from consideration and repeat the procedure using the smaller index set $J\setminus\{i,j\}$. \\
It is worth noting that each block can be analyzed independently, rendering the attack highly parallelizable.

\begin{Lemma}\label{Runtime}
    The above algorithms reconstructs the index of interest in $O(\lceil\frac{q}{\delta-1}\rceil)$ steps. 
\end{Lemma}

\begin{proof}
    The algorithm entails computing the $\mathbb{F}_q$-rank of at most $\frac{m}{2}(\lceil\frac{q}{\delta-1}\rceil+\log(\delta)+1)$ matrices, each of size at most $ns \times ns$. This computation requires a total of $\frac{m}{2}(\lceil\frac{q}{\delta-1}\rceil+\log(\delta)+1)(ns)^3$ operations over $\mathbb{F}_q$. Consequently, the algorithm's time complexity is $O\left(\lceil\frac{q}{\delta-1}\rceil\right)$. 
\end{proof}

The proposed attack significantly compromises the security guarantees claimed in \cite{Repair1}. Specifically, for the recommended parameters of the CB-cPIR scheme, the computational complexity of our attack is presented in Table \ref{AttackCom}. Note that the attack complexity is evaluated under the assumption that only one file is requested simultaneously. If, however, $\beta$ is reused multiple times, the attack complexity per file decreases further. \\

\begin{table}[h]
    \centering
    \caption{Attack complexity for the parameters proposed in \cite{Repair1} in comparison to the best-known attacks from \cite{Repair1} as  security level.}
    \begin{tabular}{@{}lrrrrr|rr@{}}
    \toprule
       \multirow{2}{*}{ q }& \multirow{2}{*}{s} & \multirow{2}{*}{v} & \multirow{2}{*}{n} & \multirow{2}{*}{k} & \multirow{2}{*}{$\delta$} & Security & Attack  \\
        &&&&&&  level & complexity\\
    \midrule
    \midrule
    32& 32& 31 & 100 & 50 & 50 & $2^{113}$& $2^1$ \\
    32 & 32 & 30 & 100 & 50& 100 & $2^{113}$& $2^1$\\
    $2^{16}$ & 12 & 10 & 100 & 50& 100 & $2^{113}$& $2^{9}$ \\
    $2^{32}-5$ & 6 & 4 & 120 & 60& 120 & $2^{128}$&  $2^{25}$\\
    $2^{32}$ & 5 & 3 & 100 & 50& 100 & $2^{96}$&$2^{25}$ \\
    $2^{61}-1$ & 6 & 2 & 100 & 50 & 200 & $2^{113}$& $2^{53}$\\
     \bottomrule
    \end{tabular}
    \label{AttackCom}
\end{table}

\begin{Remark}
    Our attack is predicated on the assumption that the database exceeds a certain size, specifically $m>ns-\delta$. With the given parameters, this translates to a database size of $400$ to $3150$, a range that is consistent with modern database scales. However, if the database is comprised of fewer files, the attack can be adapted to accommodate smaller sizes. To achieve this, rather than extracting solely the first row from each block $Q^1,\ldots,Q^m$ to construct $A$, multiple rows can be utilized. By selecting $p<\delta$ rows from each block to form $A$, the resulting matrix will comprise $pm$ rows, thereby relaxing the requirement to $mp>ns-\delta$. As a consequence, the rank of $A$ will be bounded by $ns-\delta+p$, allowing for the simultaneous evaluation of $\delta-p$ values. That way, our attack is applicable to databases of size at least $(\delta-1)m>ns-\delta$, which, for the proposed parameters, translates to a minimum of approximately 10 files in most cases. This threshold is remarkably low, rendering the attack viable for virtually all realistic database scenarios.
\end{Remark}

We successfully implemented the attack utilizing SageMath \cite{sage} on a standard laptop (Intel Core i7-1370P, 1,9 GHz, 16GB RAM). Notably, for the initial two parameter sets where $q=32$, the attack executed in a matter of seconds. However, as the value of $q$ increases, the time complexity of our attack grows in accordance with Lemma \ref{Runtime}. This highlights a limitation of our attack, as larger values of $q$ may render the attack impractical due to the increased computational requirements. Nevertheless, our results demonstrate the effectiveness of the attack for smaller values of $q$, and further optimization may be possible to improve performance for larger values.

\section{Comparison with Lattice-Based Schemes}\label{Comparison}

One natural approach to prevent our attack is to increase the field size order $q$. However, to avoid providing any information, the ratio $\frac{q}{\delta}$ must align with the desired security level. While increasing $q$ enhances security against our attack, it also significantly raises computational costs for both the server and the client, as operations in larger fields are generally more resource-intensive. Additionally, the scheme achieves higher rates when $L \gg m\delta$. As a consequence, if $q$ is already large, the files $X^i \in \mathbb{F}_q^{L \times \delta}$ become quite substantial for reasonable rates.\\

In accordance with the methodology outlined in \cite{Repair1}, we aim to conduct a comparative analysis of the communication rate of the CB-cPIR protocol with only one file being requested, incorporating the adjusted security framework, against established PIR schemes such as XPIR \cite{XPIR} and SimplePIR \cite{SimplePIR}. This comparison serves to contextualize CB-cPIR within the current landscape of state-of-the-art PIR protocols. The rate for the XPIR scheme is defined by 

\begin{align*}
R_{\text{XPIR}}(F) = \frac{F}{ms_c+F\frac{s_c}{c_p}}, 
\end{align*}
where $s_p$ and $s_c$ denote the sizes (in bits) of the plaintext and ciphertext, respectively, and  $F$ represents the bitsize of a single database file. For the XPIR scheme, we employ the parameters $(n, \log(q)) = (1024, 60)$, which provide a security level of 97 bits. As specified in \cite{Repair1}, the corresponding values of $s_c=128.000$ and $s_p=20000$ for these parameters are utilized in the calculation. For a fair comparison, we evaluate the CB-cPIR rate using the parameters $q = 2^{104}$, $s = 6$, $v = 4$, $n = 100$, and $k = 50$, which are chosen to ensure a comparable security level, taking into account all known attacks. We consider the classical CB-cPIR rate 
\begin{align*}
    R_{\text{CB-cPIR}}(F)=  \frac{F}{2ns(m\delta\log_2(q)+\frac{F}{\delta}) }, 
\end{align*}
where $F=L\delta\log_2(q)$ is the size of a single file in bits, as well as the proposed modification using a squared database with rate
\begin{align*}
    R_{\text{CB-cPIR, squared}}(F)=  \frac{F}{2ns\sqrt{m} (\log_2(q)\delta+\frac{F}{\delta}) } 
\end{align*}
(compare \cite[section 3.6.1]{Repair1}). The communication rates for a fixed number of $m=1,000$ files in the database, expressed as functions of the file size $F$ (in bits), are illustrated in Figure \ref{Comparison}. This figure, as well as the subsequent figure for comparison with SimplePIR, differs from those in \cite{Repair1}. This discrepancy is not solely due to the adjusted security level, but also because the formula presented in the original paper's figure appears to deviate from the correct expression as detailed in the text for varying file sizes in bits. Consistent with theoretical expectations, the CB-cPIR protocol employing a squared database demonstrates superior efficiency compared to its original counterpart for smaller file sizes. Conversely, for larger file sizes, the original CB-cPIR scheme exhibits enhanced performance. Notably, neither variant of CB-cPIR achieves a rate comparable to that of XPIR. \\

    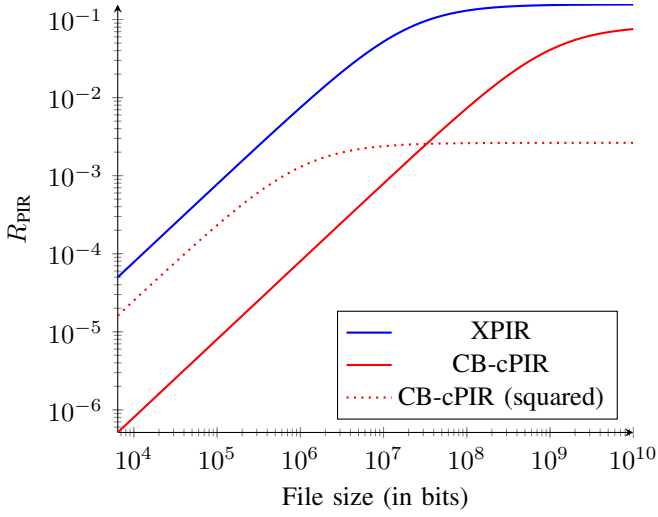
\begin{figure}
    \begin{center}
        \begin{tikzpicture}
            \begin{axis}[
            xlabel = {File size (in bits)},
            ylabel = {$R_{\text{PIR}}$},
            axis lines = left,
            legend pos = south east,
            samples =100,
            domain=6400:10000000000,
            xmode = log,
            ymode = log
            ]
            \def\m{1000}  
            \def\t{1}    
            \def\w{1}    
            \def\n{100}  
            \def\s{6}    
            \def\d{100}  
            \def \ord{104}
            
            \addplot[blue,thick] {x/(\m*128000+x*6.4)};
            \addlegendentry{XPIR}            
           \addplot [red, thick]{x/(2 * \m * \n * \s *\d*\ord + x * 2 * \n * \s/\d) };
            \addlegendentry{CB-cPIR}
            \addplot [red, dotted, thick]{x/ (2 * \m^(1/2) * \n * \s *(\d*\ord + x/\d)) };
            \addlegendentry{CB-cPIR (squared)}
            \end{axis}
        \end{tikzpicture}
    \end{center}
    \caption{Comparison of the XPIR rate for $(n,log(q))=(1024,60)$ with the rate of a comparable CB-cPIR scheme ($q = 2^{104}$, $s = 6$, $v = 4$, $n = 100$, and $k = 50$) using the original database as well as a squared version on a double-logarithmic scale. The total number of files is fixed to $m=1,000$.}
    \label{Comparison}
    \end{figure}

Similarly, we can compare the CB-cPIR scheme to SimplePIR. In the SimplePIR protocol, the user is required to download a one-time hint, which can be reused across multiple queries. Assuming the cost of downloading the hint is amortized over $t$ queries, the rate of SimplePIR can be expressed as:
\begin{align*}
    R_{\text{SimplePIR}}(F)=\frac{F\log_2(p)}{(nFt^{-1}\sqrt{m}+(F+\log_2(p))\sqrt{m})\log_2(q) },
\end{align*}
where $F$ represents the file size in bits, and $(q,p,n)$ are the scheme parameters. For our analysis, we select $(q,p,n)=(2^{32},495,1024)$, which provides a security level of $128$ bits. To achieve a comparable security level for the CB-cPIR scheme, we choose $q=2^{135}$, $s = 6$, $v = 4$, $n = 120$, and $k = 60$.\\

The resulting rates for varying file sizes, with a fixed total number of files $m=1,000$, are illustrated in Figure \ref{ComparisonSimplePIR}. The case $t=\infty$ serves as a theoretical upper bound, representing the idealized scenario where the cost of downloading the hint is ignored. This limiting case provides a benchmark for the optimal performance of SimplePIR, allowing for a direct comparison with the CB-cPIR scheme. Notably, for large file sizes, the CB-cPIR scheme may be preferred over SimplePIR. However, for smaller file sizes, the choice between the two schemes depends on how frequently the hint is reused in SimplePIR, and thus how much the cost of downloading the hint is spread across multiple queries.

    \begin{figure}
    \begin{center}
        \begin{tikzpicture}
            \begin{axis}[
            xlabel = {File size (in bits)},
            ylabel = {$R_{\text{PIR}}$},
            axis lines = left,
            legend pos = south east,
            samples =100,
            domain=6400:10000000000,
            xmode = log,
            ymode = log
            ]
            \def\m{1000}  
            \def\logp{9}   
            \def\logq{32}    
            \def\t{100}
            \def\nSimple{1024}  
            \def\n{120}
            \def\s{6}
            \def\d{120}
            \def\ord{135}

            \addplot[blue,thick,dashed] {x*\logp/((\nSimple*x*\m^(1/2)+(x+\logp)*\m^(1/2))*\logq)};
           \addlegendentry{SimplePIR ($t=1$)}
           \addplot[blue,thick,dotted] {x*\logp/((0.01*\nSimple*x*\m^(1/2)+(x+\logp)*\m^(1/2))*\logq)};
            \addlegendentry{SimplePIR ($t=100$)}
            \addplot[blue,thick] {x*\logp/(((x+\logp)*\m^(1/2))*\logq)};
            \addlegendentry{SimplePIR ($t=\infty)$}
           \addplot [red, thick]{x/(2 * \m * \n * \s *\d*\ord + x * 2 * \n * \s/\d) };
            \addlegendentry{CB-cPIR}
            \addplot [red, thick, dotted]{x/ (2 * \m^(1/2) * \n * \s *(\d*\ord + x/\d)) };
            \addlegendentry{CB-cPIR (squared)}
            \end{axis}
        \end{tikzpicture}
    \end{center}
    \caption{Comparison of the SimplePIR rate for $(q,p,n)=(2^{32},495,1024)$ with a comparable CB-cPIR scheme rate ($q = 2^{104}$, $s = 6$, $v = 4$, $n = 120$, and $k = 60$) using the original database as well as a squared version on a double-logarithmic scale. The total number of files is fixed to $m=1,000$.}
    \label{ComparisonSimplePIR}
    \end{figure}
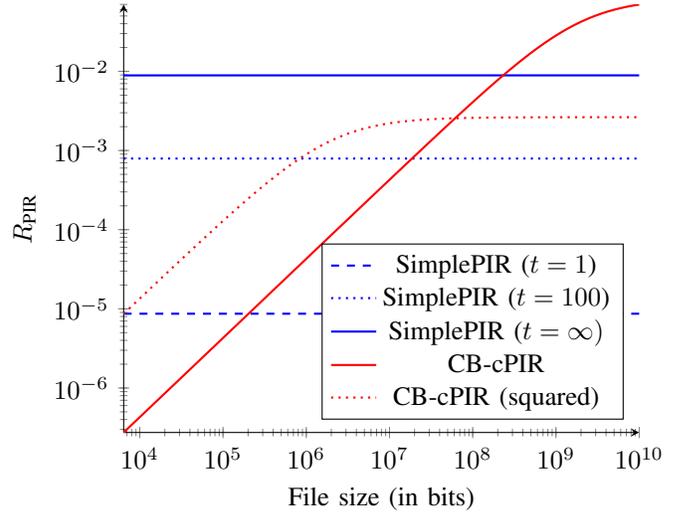

\section{Conclusion}
This paper reveals a vulnerability in the CB-cPIR scheme, demonstrating that the original parameters proposed in \cite{Repair1} no longer align with the promised security levels. While adjusting the parameters can restore the desired security level, our analysis shows that the scheme's performance is no longer competitive with state-of-the-art schemes like XPIR. However, in certain scenarios, CB-cPIR may still offer advantages over SimplePIR, depending on the specific use case and parameter choices. Our findings highlight the need for a thorough reevaluation of the CB-cPIR scheme's design and parameters to ensure its security and performance meet the expected standards. The importance of this work extends beyond the specific scheme, as it underscores the need for reliable and secure PIR schemes based on coding theory. In a landscape where most PIR schemes rely on lattice-based cryptography, coding theory-based schemes like CB-cPIR offer a vital alternative. The existence of diverse and robust PIR schemes is crucial for ensuring the long-term security and resilience of cryptographic systems.

\section*{Acknowledgment}
This work has been supported in part by funding from Agentur für Innovation in der Cybersicherheit GmbH.

\bibliographystyle{IEEEtran}

\bibliography{references}
\end{document}